\documentclass{aa}
\usepackage{graphicx}
\usepackage{txfonts}
\usepackage{natbib}
\bibpunct{(}{)}{;}{a}{}{'}
\begin{document}
\authorrunning{Pani\'c et al.}
\titlerunning{A break in the gas and dust surface density of the 
  disc around IM~Lup}
\title{A break in the gas and dust surface density of the disc around
  the T~Tauri star IM~Lup}
\author{O. Pani\'c\inst{1} \and M. R. Hogerheijde\inst{1} \and
  D. Wilner\inst{2} \and C. Qi\inst{2}}
\institute{Leiden Observatory, Leiden University, P.O.Box 9513, 2300
  RA, Leiden, The Netherlands
\and
  Harvard-Smithsonian Center for Astrophysics, 60 Garden Street,
  Cambridge, MA 02138, USA
}
\date{Received  ; Accepted }

\abstract
{}
{We study the distribution and physical properties of molecular gas in
  the disc around the T Tauri star IM Lup on scales close to 200~AU. We
  investigate how well the gas and dust distributions compare and work
  towards a unified disc model that can explain both gas and dust
  emission.}
{$^{12}$CO, $^{13}$CO, and C$^{18}$O  $J$=2--1 line emission, as well as the dust
  continuum at 1.3~mm, is observed at 1$\farcs$8 resolution towards IM
  Lup using the Submillimeter Array. A detailed disc model based on
  the dust emission is tested against these observations with the aid of a
  molecular excitation and radiative transfer code. Apparent
  discrepancies between the gas and dust distribution are investigated by
  adopting simple modifications to the existing model.}
{The disc is seen at an inclination of 54$\degr$ $\pm $3$\degr$ and
  is in Keplerian rotation around a 0.8--1.6~M$_\odot$ star. The
  outer disc radius traced by molecular gas emission is 900~AU, while
  the dust continuum emission and scattered light images limit the amount of dust present beyond 400~AU and are consistent with the existing model that assumes a 400~AU radius.
  Our observations require a drastic density decrease close to 400~AU 
  with the vertical gas column density at 900~AU in the range of $5\times 10^{20}$--$10^{22}$~cm$^{-2}$. 
  We derive a gas-to-dust mass ratio of 100 or higher in disc regions beyond
  400~AU. Within 400~AU from the star our observations are consistent
  with a gas-to-dust ratio of 100 but other values are not ruled out.}
{}

\maketitle

\section{Introduction}\label{s:intro}

Low-mass star formation is commonly accompanied by the formation of a circumstellar disc. 
The disc inherits the angular momentum and composition of the star's parent cloud, and is shaped 
by the accretion and other physical processes within the disc during the evolution that may 
result in a planetary system. 
Over the past two decades observations of circumstellar discs at millimetre
wavelengths have been established as powerful probes of the bulk of
the cold molecular gas and dust. Spatially resolved observations of
the molecular gas with (sub) millimetre interferometers constrain the
disc size and inclination, the total amount of gas, its radial and
vertical structure, and the gas kinematics \citep[e.g.,][]{guilloteau,
  dartois, qi, isella, pietu, panic}.  In parallel, continuum
observations of the dust at near-infrared to millimetre wavelengths
provide the disc spectral energy distribution (SED), that through
modelling \citep[e.g.,][]{chiang,dullemond,dalessio} yields the disc's
density and temperature structure from the disc inner radius to a few
hundred AU from the star. Studies of the gas through
spatially resolved molecular line observations using results from the
SED modelling \citep[e.g.,][]{raman,panic} offer the means of addressing the
gas-to-dust ratio, differences between the radial and vertical
distributions of the gas and the dust, and the thermal coupling between
the gas and the dust in the upper disc layers exposed to the stellar
radiation \citep[e.g.,][]{jonkheid}. Recent papers have suggested
different disc sizes for the dust and the gas (e.g., HD 163296,
\citealt{isella}), which may 
be explained by sensitivity effects in discs with tapered outer density
profiles \citep{mhughes}. Here, we present the results of a combined
study using spatially resolved molecular-line observations and SED
modelling of the disc around the low-mass pre-main-sequence star
\object{IM Lup}.

Most pre-main-sequence stars with discs studied so far in detail are
located in the nearby star-forming region of Taurus-Aurigae,
accessible for the millimetre facilities in the northern
hemisphere. Much less is known about discs in other star-forming
regions such as Lupus, Ophiuchus or Chamaeleon. It is important to
compare discs between different regions, to investigate if and how
different star-forming environments lead to differences in disc
properties and the subsequent planetary systems that may result.
IM Lup is a pre-main-sequence star located in the Lupus~II cloud for
which \citet{wichmann} report a distance of 190$\pm$27~pc using
Hipparcos parallaxes. From its M0 spectral type and estimated
bolometric luminosity of 1.3$\pm$0.3~L$_\odot$, \citet{hughes} derive a
mass of 0.4~M$_\odot$ and an age of 0.6~Myr using evolutionary tracks
from \citet{swenson94}, or 0.3~M$_{\odot}$ and 0.1~Myr adopting the
tracks of \citet{d'antona94}. In \citet{pinte}, a much higher value of 1~M$_{\odot}$ is derived using tracks of \citet{baraffe}.

IM Lup is surrounded by a disc detected in scattered light with the
Hubble Space Telescope \citep{pinte}
and in the
CO $J$=3--2 line with the James Clerk Maxwell Telescope by
\citet{vankempen}. 
\citet{lommen} conclude that grain growth up
to millimetre sizes has occured from continuum measurements at 1.4 and
3.3~mm. Recently, \citet{pinte} present a detailed model for the disc
around IM Lup based on the full SED,
scattered light images at multiple wavelengths from the Hubble Space
Telescope, near- and mid-infrared spectroscopy from the Spitzer Space
Telescope, and continuum imaging at 1.3~mm with the Submillimeter
Array. They conclude that the disc is relatively massive, $M\approx
0.1$~M$_\odot$ with an uncertainty by a factor of a few, has an outer dust radius 
not greater than $\approx$400~AU set by the dark lane and lower reflection lobe seen in
the scattered light images, and has a surface density $\Sigma(R)$
proportional to $R^{-1}$ constrained by the 1.3~mm data. Furthermore,
they find evidence for vertical settling of larger grains by comparing
the short-wavelength scattering properties to the grain-size
constraints obtained at 1.4 and 3.3~mm by \citet{lommen}.

In this work, we present (Sect.~\ref{s:obs}) spatially resolved
observations of the disc around IM Lup in $^{12}$CO, $^{13}$CO and
C$^{18}$O $J$=2--1 line emission, together with 1.3~mm dust continuum
data, obtained with the Submillimeter Array\footnote{The Submillimeter
  Array is a joint project between the Smithsonian Astrophysical
  Observatory and the Academia Sinica Institute of Astronomy and
  Astrophysics and is funded by the Smithsonian Institution and the
  Academia Sinica.}  (SMA). Our results (Sect.~\ref{s:results}) show that
the gas disc extends to a radius of 900~AU, more than twice the size
inferred by \citet{pinte}. A detailed comparison (Sect.~\ref{s:cfpinte})
to the model of \citet{pinte} suggests a significant break in the
surface density of both the gas and the dust around 400~AU, and we
discuss possible explanations. We summarise our main conclusions in
Sect.~\ref{s:conclusions}.


\section{Observations}\label{s:obs}

IM Lup was observed with the SMA on 2006 May 21 in a 8.6-hour
track, with a 4.3 hours on-source integration time. The coordinates of the phase centre are RA$=$
15$^\mathrm{h}$56$^\mathrm{m}$09$\fs$17 and Dec$=-$37$\degr$56$\arcmin$06$\farcs$40 (J2000). Eight
antennas were included in an extended configuration providing a range
of projected baselines of 7 to 140 meters. The primary beam
half-power width is $55\farcs0$. The SMA receivers operate in a double-sideband (DSB) mode with an intermediate frequency band of 4--6~GHz which is sent over fiber optic transmission lines to 24 ``overlapping'' digital correlator chunks covering a 2~GHz spectral window in each sideband. The DSB system temperatures ranged from 90 to 150~K.
The correlator was configured to include the $^{12}$CO $J$=2--1 line
(230.5380000~GHz) in the upper sideband and the $^{13}$CO 2--1
(220.3986765~GHz) and C$^{18}$O 2--1 line (219.5603568~GHz) in the
lower sideband. Each of the three lines was recorded in a spectral
band consisting of 512 channels with 0.2~MHz spacing
($\sim$0.26~km~s$^{-1}$). Simultaneously to the line observations,
the 1.3~mm dust continuum was recorded over a bandwidth of 1.8~GHz.

The data were calibrated and edited with the IDL-based \texttt{MIR~}software package (\footnote{http://www.cfa.harvard.edu/$\sim$cqi/mircook.html}). The bandpass response was determined from Jupiter, Callisto and 3C273. After the passband calibration, broadband continuum in each sideband was generated by averaging the central 82~MHz in all line-free chunks. Complex gain calibration was performed using the quasar J1626$-$298. 
The absolute flux scale was set using observations of Callisto. 
Subsequent 
data reduction and image analysis was carried out with the Miriad 
software package \citep{sault}.
The visibilities were Fourier transformed with natural weighting,
resulting in a synthesized beam of $1\farcs8\times1\farcs2$ at a
position angle of $0.2^{\circ}$. 1~Jy/beam corresponds
to 15.9 K. The r.m.s noise level is
125, 94 and 102~mJy~beam$^{-1}$ per channel respectively for the $^{12}$CO, $^{13}$CO
and C$^{18}$O spectral line data and 4~mJy~beam$^{-1}$ for the continuum data.


\section{Results}\label{s:results}

\subsection{Dust Continuum}\label{s:continuum}

\begin{figure}
\centering
\includegraphics[angle=0.,width=8.5cm]{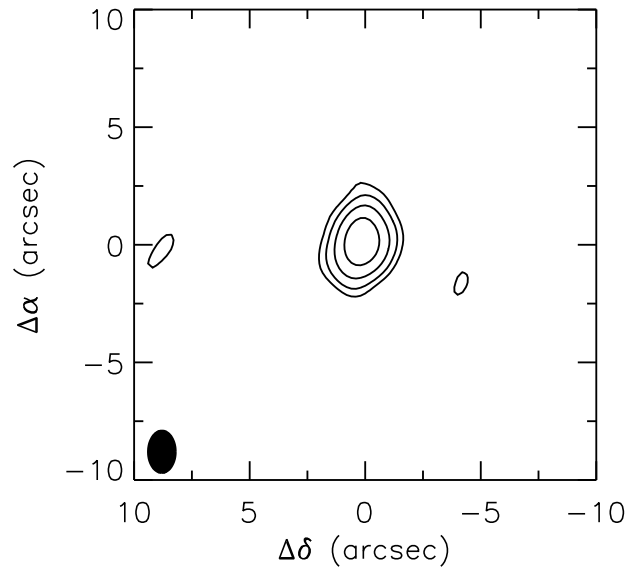}
\caption
      {Dust continuum image at 1.3~mm. The contours are at (2, 4, 8, 16)$\times$3.67~mJy~beam$^{-1}$ (2, 4, 8, 16 sigma) levels. The filled ellipse in the lower left corner shows the synthesized beam.}
 \label{dust} 
\end{figure}

\begin{figure}
\includegraphics[angle=-90,width=9cm]{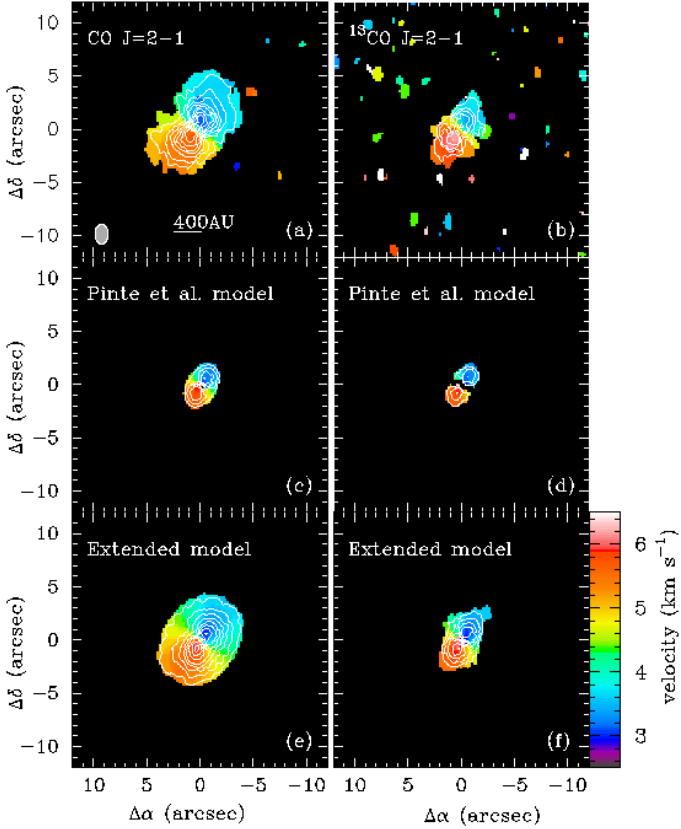}
\caption
      {\textit{(a),(b):} First moment maps in the $^{12}$CO and $^{13}$CO
         $J$=2--1 lines, from 1.9~kms$^{-1}$ to 6.9~kms$^{-1}$ observed towards IM Lup. These maps
        are created using the Miriad task `moment' with clip levels of
        0.5 and 0.3~Jy respectively. The integrated emission of $^{12}$CO $J$=2--1 is shown in
        contours of 1, 2, 3, ...$\times$500~mJy, and that of $^{13}$CO $J$=2--1
        with 1, 2, 3, ...$\times$160~mJy contours. \textit{(c), (d):} First moment and integrated emission maps calculated using \citeauthor{pinte} model and same clip level, velocity range and contour levels as in (a) and (b). \textit{(e), (f):} First moment and integrated emission maps calculated using extended disc model (described in Sect.~\ref{s:model}) with model parameters $Sigma_{400}=$2$\times$10$^{21}$~cm$^{-2}$ and $p=$1. The clip level, velocity range and contour levels are as in (a), (b), (c) and (d).}
 \label{mom} 
\end{figure}

Figure~\ref{dust} shows the 1.3~mm continuum emission observed toward
IM Lup, previously reported in \citet{pinte}. The emission is centered on RA$=$15$^\mathrm{d}$56$^\mathrm{m}$09$\fs$20, Dec$=-$37$\degr$56$\arcmin$06$\farcs$5 (J2000), offset by +0$\farcs$4 in right ascension and by $-$0$\farcs$1 in declination
  from the pointing center. We adopt the peak of the
  continuum emission as the position of IM Lup. The peak intensity of
  the continuum emission is $104\pm 4 $~mJy~beam$^{-1}$ and the total
  flux 176$\pm$4~mJy. The emission intensity is fit to the precision of one sigma by an elliptical
  Gaussian, yielding a source FWHM size of 1$\farcs$52$\pm$0$\farcs$15 $\times$ 1$\farcs$06$\pm$0$\farcs$15
  and a position angle of $-35{\fdg}5\pm 4\fdg0$ deconvolved with the
  synthesized beam. This position angle, and the inclination in the range of 33$\degr$--53$\degr$ 
  suggested by the deconvolved aspect ratio, agree well with the values obtained by \citet{pinte} 
  of, respectively, $-$37$\degr\pm$5$\degr$ and 45$\degr$--53$\degr$ from scattered light imaging.

A fit to the 1.3~mm visibilities done in \citet{pinte} provides a rough disc mass estimate of 0.1~M$_\odot$, with an uncertainty of a factor of few, dominated by the adopted dust emissivity and gas-to-dust mass ratio in the model.


\subsection{Molecular Lines}\label{s:lines}

Emission of $^{12}$CO and $^{13}$CO $J$=2--1 was detected toward
IM~Lup, and an upper limit on C$^{18}$O 2--1
obtained. Figure~\ref{mom} shows the zero moment (integrated emission,
contours) and first moment (velocity centroid, colour scale) of the
$^{12}$CO and $^{13}$CO emission from IM Lup. Significantly detected
$^{12}$CO emission extends to 5$\arcsec$ from the star (roughly 900~AU for IM Lup).
This is more than double the size inferred from
the dust continuum image, and Sect.~\ref{s:discussion} discusses if
this is due to different sensitivity in these two tracers or if the
gas disc indeed extends further than the dust disc. The aspect ratio
(5/3), suggesting an inclination of $53\degr \pm 4\degr$, and
orientation PA=$-$36$\pm$5$\degr$ of the CO disc agree with the continuum image
(Sect.~\ref{s:continuum}) and scattered light imaging results \citep{pinte}.

\begin{figure}
\includegraphics[angle=0,width=7cm]{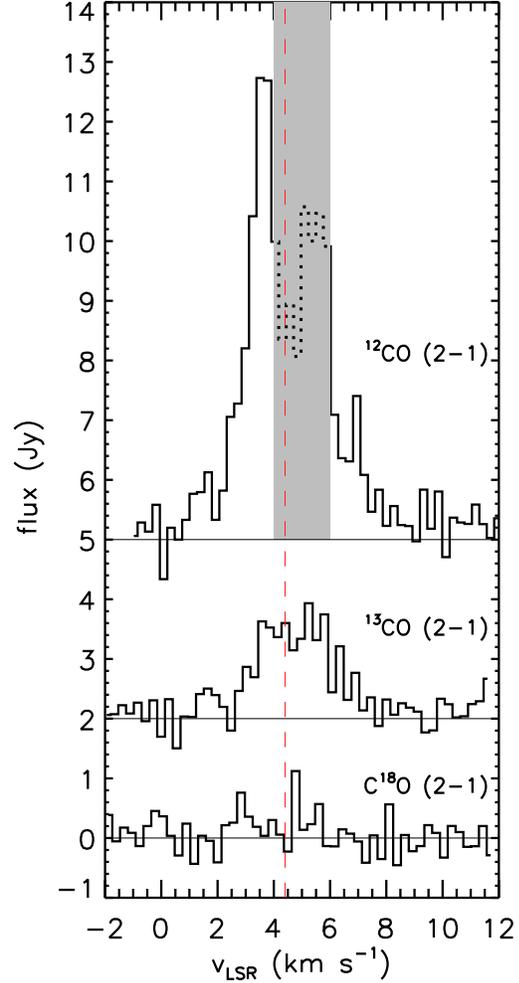}
\caption
      {$^{12}$CO, $^{13}$CO, and C$^{18}$O $J$=2--1 line spectra
        summed over $8\arcsec\times8\arcsec$ regions centered on the
        location of IM Lup. The $^{13}$CO and $^{12}$CO spectra are
        shifted upward by 2 and 5~Jy, respectively. The dashed red
        line shows the line centre at $v_{LSR}$=4.4~km~s$^{-1}$. The
        grey zone indicates the range from 4 to 6~km~s$^{-1}$ where
        the $^{12}$CO line is significantly affected by the foreground
        absorption; the corresponding part of the $^{12}$CO spectrum
        is plotted with a dotted line.}
\label{spec}
\end{figure}

\begin{figure*}[!htp]
\centering
\includegraphics[angle=0,width=19cm]{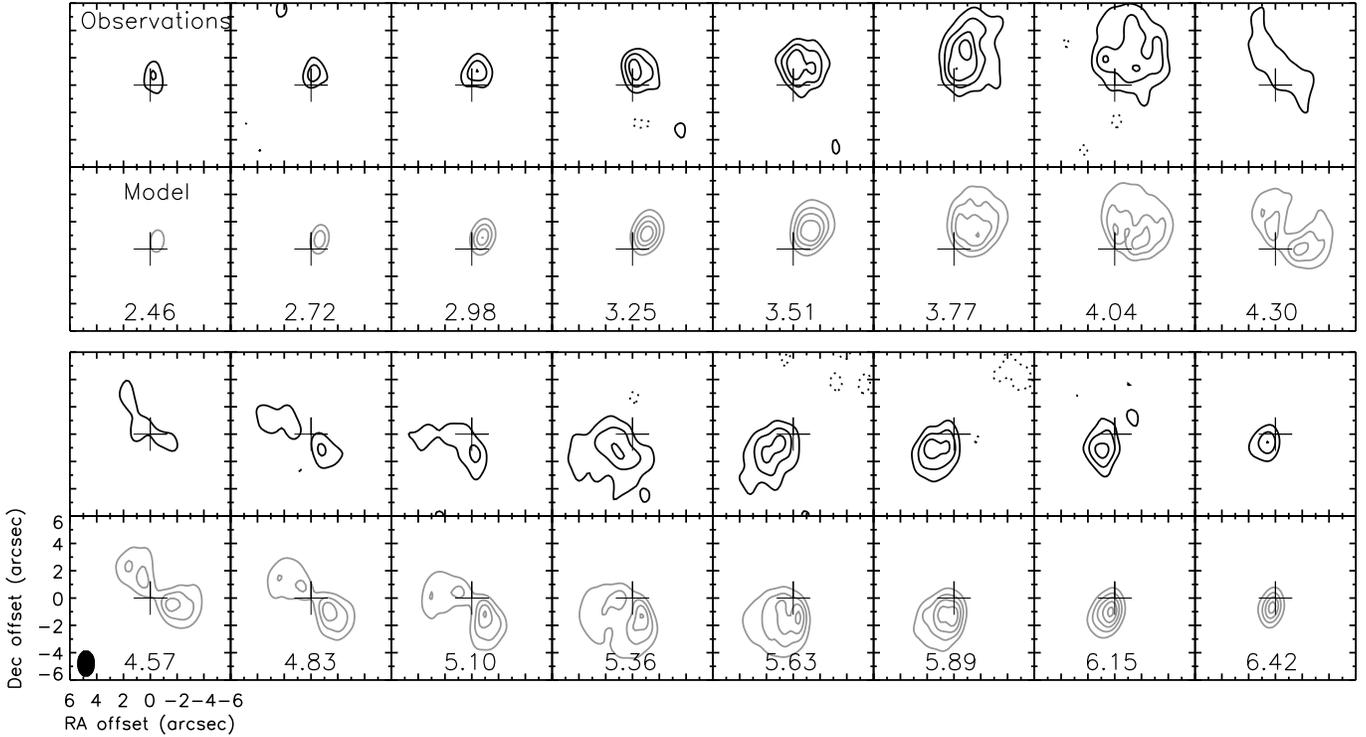}
\caption
        {The black contours show the observed $^{12}$CO $J$=2--1 emission in the velocity range from 
	2.46 to 6.42~km~s$^{-1}$. Alongside the observations, the panels with grey contours show the
	calculated emission from the extended disc model described in
        Sect.~\ref{s:model}, with parameters $\Sigma_{400}=2\times10^{21}$cm$^{-2}$ and $p=$1. 
	Labels indicate the velocity of each channel. The lower left corner of bottom-left panel shows the 
	size and position angle of the synthesized beam. The contour levels are $-$1, 1, 2, 3,
	4$\times$400~mJy~beam$^{-1}$ ($\sim$3$\sigma$) in all panels.}
 \label{maps12} 
\end{figure*}

\begin{figure*}
\includegraphics[angle=0,width=19cm]{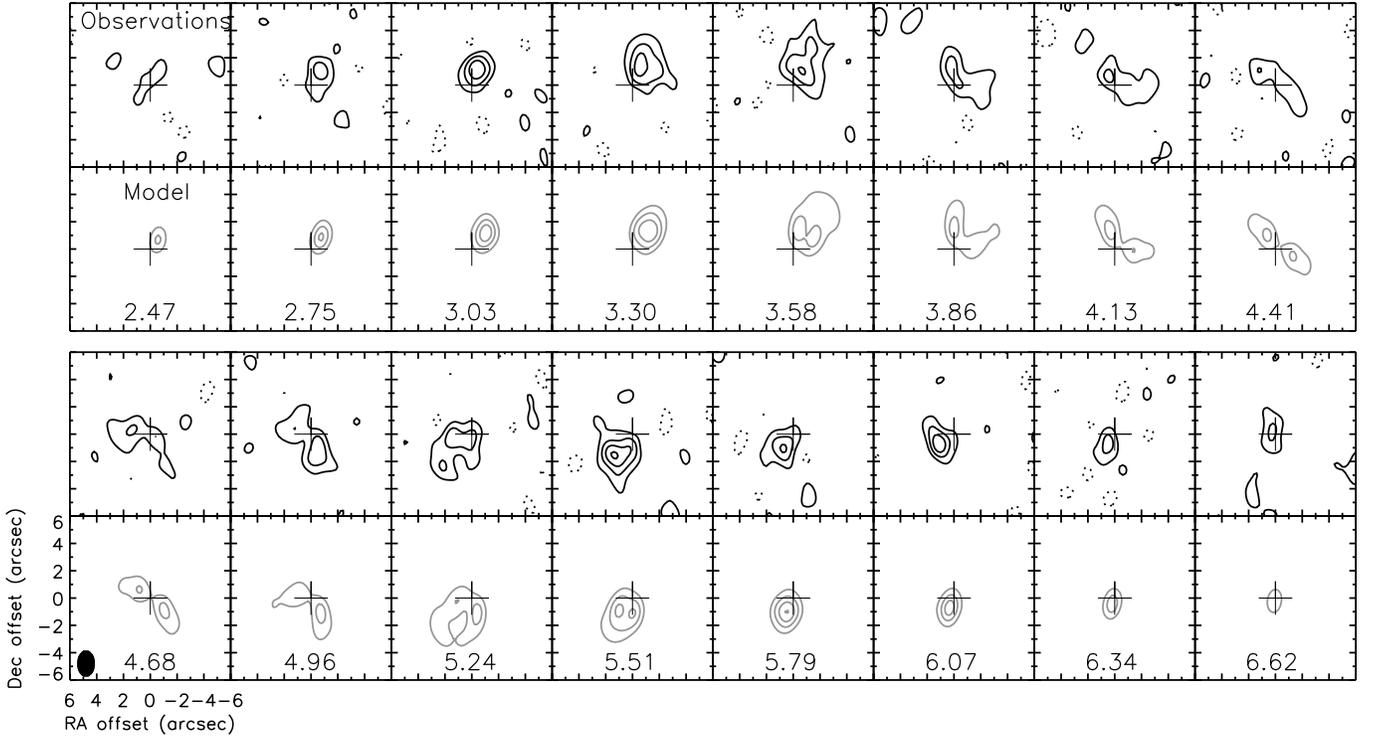}
\caption
      	{Channel maps of the observed $^{13}$CO $J$=2--1 emission at the velocities where the line is
        detected are shown in black contours. For comparison, the line emission calculated from our 
	extended disc model described in Sect.~\ref{s:model} is shown in grey contours. The model 
	parameters are $\Sigma_{400}=2\times10^{21}$cm$^{-2}$ and $p=$1. Labels
        indicate the velocity of each channel. The lower left corner of bottom-left panel shows the size
        and position angle of the synthesized beam. The contour levels
        are $-$1, 1, 2, 3, 4$\times$200~mJy~beam$^{-1}$ ($\sim$2$\sigma$) in all panels.}
\label{mapsA13}
\end{figure*}

\begin{figure}
\includegraphics[angle=0,width=9cm]{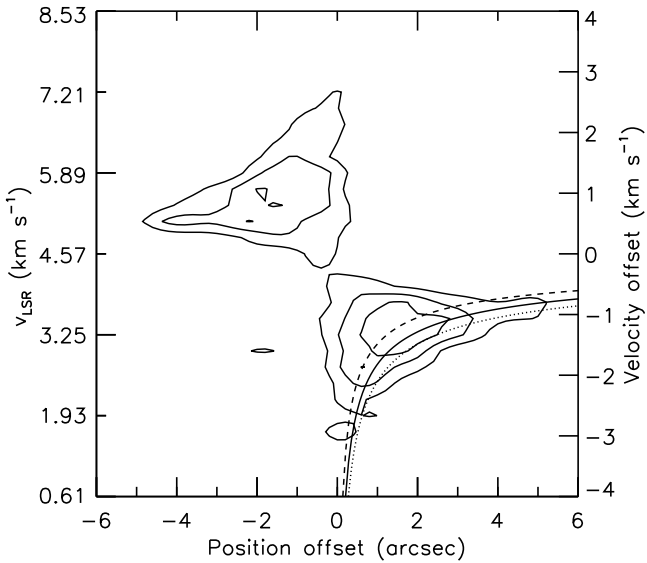}
\caption
      {Position-velocity diagram of the $^{12}$CO 2--1 line emission
        along the disc's major axis. Contour levels are (1, 2, 3,
        ...)$\times$400~mJy ($\sim$3$\sigma$). For comparison, the thick solid line
        corresponds to Keplerian rotation around a 1.2~M$_\odot$ star; dashed and
        dotted lines correspond to the stellar masses of 0.8 and of
        1.6~M$_\odot$, respectively.}
\label{vlplt}
\end{figure}

The first moment images of Fig.~\ref{mom} show velocity patterns
indicative of a rotating disc inclined with respect to the line of sight. This is also
seen in Fig.~\ref{spec}, which presents the $^{12}$CO, $^{13}$CO, and
C$^{18}$O spectra averaged over $8'' \times 8''$ boxes around IM
Lup. The $^{12}$CO and $^{13}$CO lines are double-peaked and centered on ${\rm v_{LSR}}=$4.4$\pm$0.3~km~s$^{-1}$. Figures \ref{maps12} and \ref{mapsA13} show the channel maps of the $^{12}$CO and $^{13}$CO
emission,
respectively, revealing the same velocity pattern also seen from the
first-moment maps and the spectra. The Keplerian nature of the
velocity pattern is most clearly revealed by Fig.~\ref{vlplt}, which
shows the position-velocity diagram of the $^{12}$CO emission along
the major axis of the disc. In Section \ref{s:discussion}, we derive a stellar mass of 1.2~M$_\odot$, and, as an illustration,
the rotation curves for stellar masses of 0.8, 1.2, and 1.6~M$_\odot$
are plotted in Fig.~\ref{vlplt}.

Using single-dish $^{12}$CO 3--2 observations, \citet{vankempen} first
identified molecular gas directly associated with IM~Lup, but they
also conclude that the $v_{LSR}$-range of 4 to 6~km~s$^{-1}$ is
dominated by gas distributed over a larger ($>30''$) scale. In our
$^{12}$CO 2--1 data this same $v_{LSR}$-range is also likely affected:
where the single-dish $^{12}$CO 3--2 spectrum from
\citeauthor{vankempen} shows excess emission over
$v_{LSR}=$4--6~km~s$^{-1}$, the red peak of our $^{12}$CO 2--1
spectrum, which lies in this $v_{LSR}$-range, is weaker than the blue
peak at $+3.5$~km~s$^{-1}$. We suspect that absorption by the same
foreground layer identified by \citeauthor{vankempen} is resonsible
for this decrement, while its emission is filtered out by the
interferometer. The $^{13}$CO 2--1 spectrum is symmetric, suggesting
that the foreground layer is optically thin in this line.

\begin{figure*}
\begin{center}
       \includegraphics[width=9.cm]{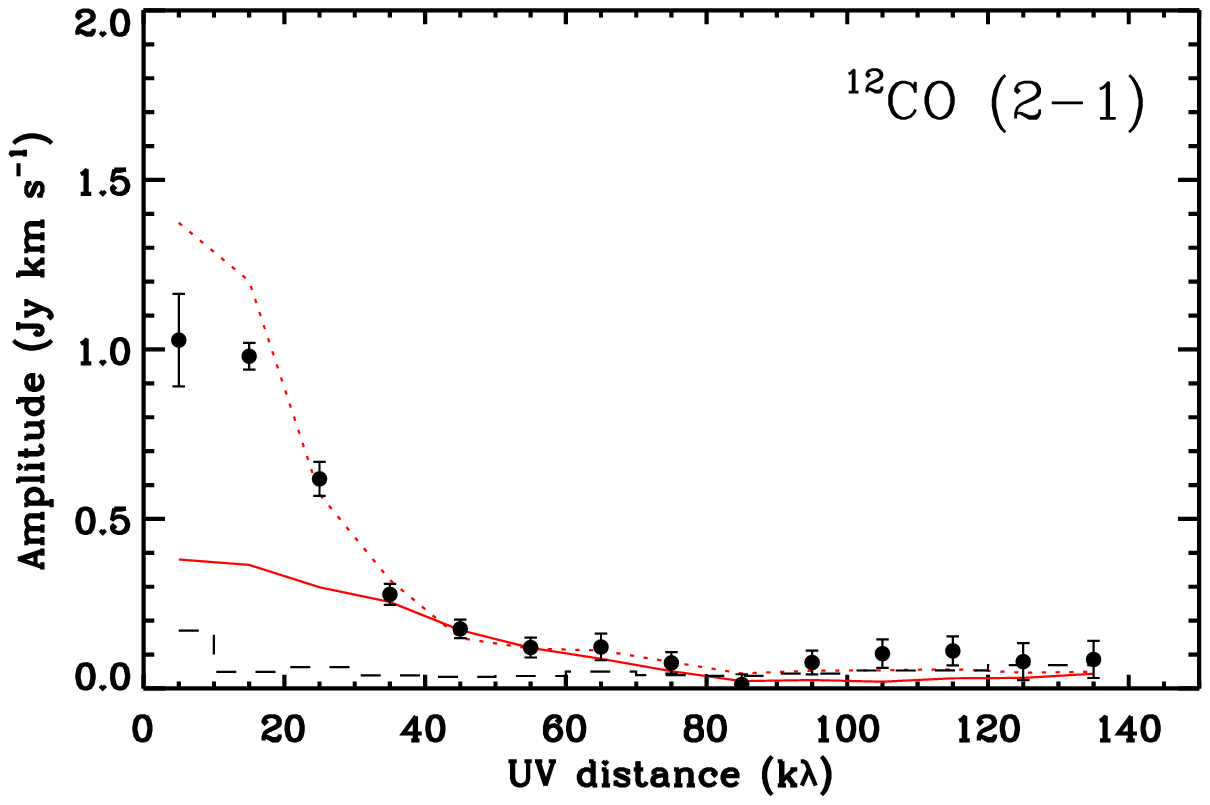}
     \includegraphics[width=9.cm]{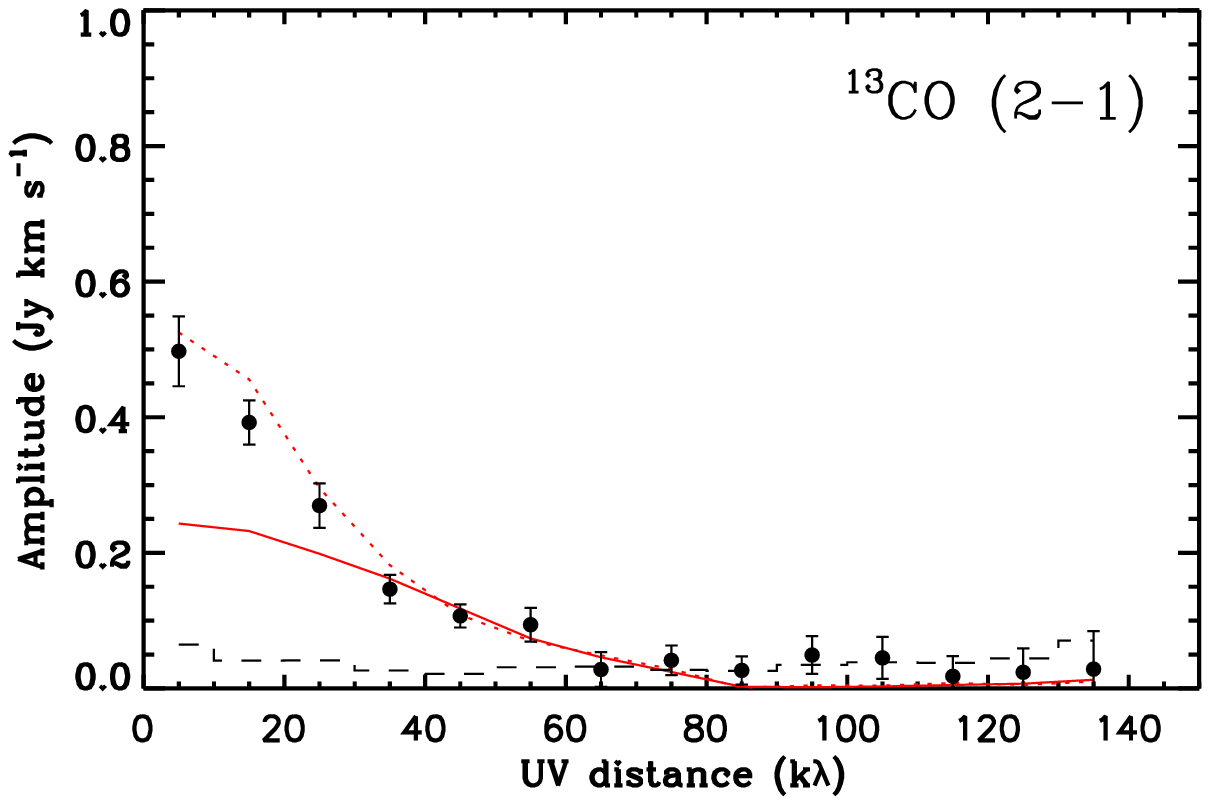}
\caption{The left and the right panel show vector-averaged $^{12}$CO and $^{13}$CO line flux (black symbols),
	respectively. Dashed black lines represent the zero-signal expectation value of our line visibility
	data. The calculated visibilities based on \citeauthor{pinte} model (full red line) and 
	our extended disc model described in Sect.~\ref{s:model} (dotted red line) are shown for comparison.
	Our model parameters are
	$\Sigma_{400}=2\times10^{21}$~cm$^{-2}$ and $p=$1. The $^{12}$CO flux is integrated
	over the 0.8-4.0~kms$^{-1}$ range and $^{13}$CO over 2.5-6.9~kms$^{-1}$, covering the full line
	width.}
\label{uvco}
\end{center}
\end{figure*}

\begin{figure}
\centering
\includegraphics[angle=0.,width=9cm]{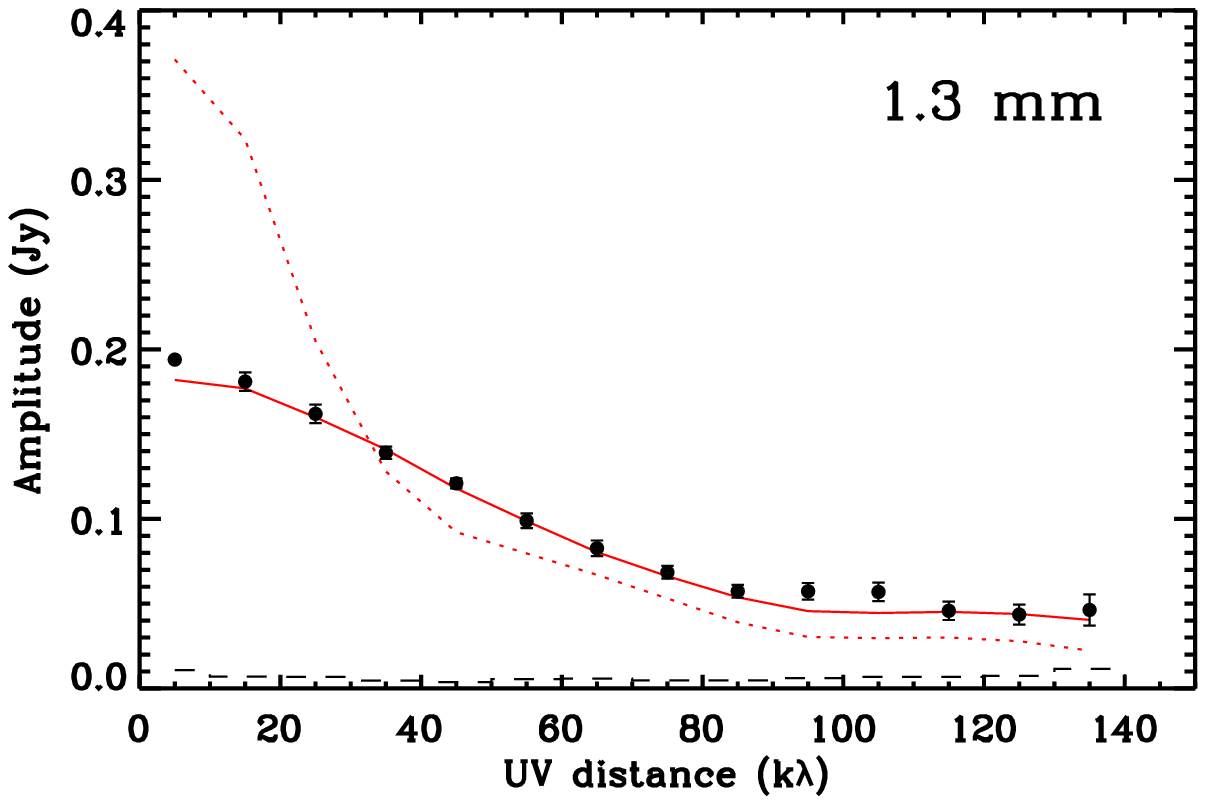}
\caption{Vector-averaged continuum flux as a function of projected
  baseline length (black symbols). Error bars show the variance within
  each annular bin. The dashed histogram shows the zero-signal
  expectation value. The full red line shows the continuum flux calculated 
  from \citeauthor{pinte} model. The dotted red line corresponds to the extended disc model with $\Sigma_{400}=2\times$10$^{21}$~cm$^{-2}$ and $p=$1 (see Sect.~\ref{s:model} for model description).}
 \label{uvamp} 
\end{figure}

The spatial extent of the line emission is further explored in
Fig.~\ref{uvco} which plots the $^{12}$CO and $^{13}$CO  $J$=2--1
vector-averaged line fluxes against projected baseline length. The
$^{12}$CO flux is integrated from 2.5 to 4.0~kms$^{-1}$ to avoid the
range where foreground absorption affects the line. The $^{13}$CO flux
does not suffer from absorption and is integrated over its full extent
from 2.5 to 6.9~kms$^{-1}$. Comparing the curves of Fig.~\ref{uvco} to
those of the continuum flux versus baseline lengths (Fig.~\ref{uvamp}) it
is clear that the line flux is much more dominated by short spacings ($<$40~k${\rm \lambda}$). This profile may indicate the presence of a larger structural component (outer disc or envelope), combined with the disc emission \citep[See][ Fig.~2]{jorgensen}. We explore disc structure beyond 400~AU in the following section.

\section{Discussion}\label{s:discussion}

The results of the previous section show that IM~Lup is surrounded by
a gaseous disc in (roughly) Keplerian rotation. The gas disc has a
radius of 900~AU, and its surface density may have a break around a
radius of 400~AU. In contrast, the size of the dust disc is
constrained to a radius of 400~AU by our continuum data and the
modelling of \citet{pinte}. This section explores if the gas and dust
have the same spatial distribution (in which case different
sensitivity levels need to explain the apparent difference in size) or
if the gas and dust are differently distributed radially. First we
investigate whether the model of \citet{pinte} can explain the
molecular line observations (Sect.~\ref{s:cfpinte}). After we conclude
that this is not the case, we construct new models for the gas disc
(Sect.~\ref{s:model}) describing their best-fit parameters,
and compare them to the dust disc
(Sect.~\ref{s:cfdust}).


\subsection{Molecular-line emission from the dust-disc model}\label{s:cfpinte}

Recently, \citet{pinte} present a detailed model for the disc
around IM Lup based on the full SED,
scattered light images at multiple wavelengths from the Hubble Space
Telescope, near- and mid-infrared spectroscopy from the Spitzer Space
Telescope, and continuum imaging at 1.3~mm with the Submillimeter
Array.

Based on the two-dimensional density and temperature structure of the Pinte et al. model, with
$M=0.1$~M$_\odot$, $R_{out}=400$~AU, and $i=50\degr$, we
calculate the resulting line intensity of the $^{12}$CO and $^{13}$CO
 $J$=2--1 lines. To generate the input model for the molecular
excitation calculations, we adopt a gas-to-dust mass ratio of 100 and
molecular abundances typical for the dense interstellar medium
\citep{frerking,wilson}: a $^{12}$CO abundance with respect to H$_2$
of $10^{-4}$ and a $^{12}$CO/$^{13}$CO abundance ratio of 77. No
freeze-out or photodissociation of CO is included. The velocity of the
material in the disc is described by Keplerian rotation around a
1.0~M$_\odot$ star plus a Gaussian microturbulent velocity field with a
FWHM of 0.16~km~s$^{-1}$; the exact value of the latter parameter has little effect on the
results. Using the molecular excitation and radiative transfer code
RATRAN \citep{hogerheijde} and CO-H$_2$ collision rates from the
Leiden Atomic and Molecular Database
(LAMBDA\footnote{http://www.strw.leidenuniv.nl/$\sim$moldata/};
\citealt{schoier}) we calculate the sky brightness distribution of
the disc in the $^{12}$CO and $^{13}$CO  $J$=2--1 lines for its distance of
190~pc. From the resulting image cube, synthetic visibilities
corresponding to the actual $(u,v)$ positions of our SMA data were
produced using the MIRIAD package \citep{sault}. Subsequent Fourier
transforming, cleaning, and image restoration was performed with the
same software.

Figure~\ref{mom} compares the zeroth-moment (integrated intensity;
contours) and first-moment (velocity-integrated intensity; colour scale) maps of
the resulting synthetic images to the observations. Clearly, the
\citeauthor{pinte} model produces $^{12}$CO and $^{13}$CO 2--1
emission with spatial extents and intensities too small by a factor close to two. In Fig.~\ref{uvco} it is clear that the \citeauthor{pinte} model fails to reproduce the $^{12}$CO and $^{13}$CO line fluxes at short projected baseline lengths, but is consistent with the observations longward of 40~k${\rm \lambda}$ that correspond to spatial scales $\leq$500~AU. 
Our comparison with \citeauthor{pinte} model thus suggests that the gas extends much further
than 400~AU from the star. 

The observed 1.3~mm continuum emission traces the extent of larger dust particles (up to millimetre in size). \citet{pinte} show that their 400~AU model reproduces these observations well. In
Sect.~\ref{s:cfdust} we explore to what level larger particles can be present
outside 400~AU.


\subsection{Extending the gas disc beyond 400 AU}\label{s:model}

As mentioned in Sect.~\ref{s:lines}, the CO line flux as function of
projected baseline length suggests a possible break in the emission
around 40~k$\lambda$ (Fig.~\ref{uvco}). Results of Sect.~\ref{s:cfpinte} show that the
\citeauthor{pinte} model, while providing a good description of line fluxes at small spatial scales (baselines$>40\rm{k}\lambda$), requires a more extended component to match the observed line fluxes (baselines$<40\rm{k}\lambda$). In this section we extend the \citeauthor{pinte} model by simple radial power laws for the gas surface density and temperature, and place limits on the gas
column densities in the region between 400 and 900~AU.

\begin{table*}
 \caption{Model parameters}
\label{tab}
\centering
\begin{tabular}{c | c | c }
 Parameter & $R<$400~AU & 400$\leq R\leq$800~AU \\
\hline\hline
 Surface density & 
   $\Sigma \propto R^{-1}$, see \citet{pinte} & 
   $\Sigma_{400}\,(R/400\,{\rm AU})^{-p}$ 
   with $\Sigma_{400}\leq$0.9~g~cm$^{-2}$, $p\geq 0$ \\
 Gas-to-dust mass ratio & 100 & 100 \\
 Gas temperature structure & 
   $f\,T_{\rm dust}$ with $1 \leq f \leq 2$ & 
   $T_{400}\,(R/400~{\rm AU})^{-q}$, $T_{400}=30$~K, $q=0.5$ \\
 Vertical structure  & 
   see \citet{pinte} & 
   $T(z)={\rm constant}$, $\rho(\rm{z})={\rm constant}$, $z_{max}=100$~AU \\
 ${\rm [CO]/[H_2] }$ & $10^{-4}$ & $10^{-4}$ \\
 ${\rm [^{12}CO]/[^{13}CO]}$ & 77 & 77 \\
 $M_\star$ & 1.2 M$_\odot$ & 1.2 M$_\odot$ \\
 Inclination & 50$\degr$ & 50$\degr$ \\
 FWHM microtubulence & 0.16 km~s$^{-1}$ & 0.16 km~s$^{-1}$ \\
\hline
\end{tabular}
\end{table*}

Table~\ref{tab} lists the model parameters. For radii smaller than
$400$~AU, the radial and vertical density distribution of the
material follows the \citeauthor{pinte} model. As in Sect.~\ref{s:cfpinte} we adopt `standard' values of gas-to-dust mass ratio and molecular abundances, and a Gaussian microturbulent
velocity field with equivalent line width of 0.16~km~s$^{-1}$. Unlike the calculations of
Sect.~\ref{s:cfpinte} we add as free parameters the stellar mass $M_\star$
and the gas kinetic temperature. For the latter, we follow the
two-dimensional structure prescribed by \citeauthor{pinte}, but scale
the temperatures upward by a factor $f$ with $1\leq f\leq 2$. This
corresponds to a decoupling of the gas and dust temperatures, as may
be expected at the significant height above the midplane where the
$^{12}$CO and $^{13}$CO lines originate \citep[see, e.g.,][]{qi1,jonkheid}. Because the highly red- and blue-shifted line emission (line wings)
comes from regions closer to the star than 400~AU and is optically thick, factor f is determined by the observed fluxes in the line wings. The molecular excitation and synthetic line data
are produced in the same way as described in Sect.~\ref{s:cfpinte}.

Outside 400~AU we extend the disc to
900~AU, as suggested by the observed $^{12}$CO image of Fig.~\ref{mom},
by simple radial power laws for the surface density and temperature,
$\Sigma$=$\Sigma_{400}\,({\rm R}/400\,{\rm AU})^{-p}$ and $T=T_{400}\,
(R/400~{\rm AU})^{-q}$. At 400~AU, the surface density is
$\Sigma_{400}$ and the temperature is $T_{400}$; the parameter $p$ is assumed to be $\ge 0$. 
To limit the number of free parameters,
we set $T_{400}=30$~K and $q=0.5$; we assume that the disc is
vertically isothermal and that the $^{12}$CO abundance is 10$^{-4}$, constant throughout the disc. At $R>$400~AU, the disc thickness is set to $z_{\rm max}$=100~AU and the density $\rho(R,z)$=$\Sigma(R)/z_{\rm max}$ is vertically constant.
For our free parameters $\Sigma_{400}$ and $p$, we assume that
$\Sigma_{400}\le$0.9~g~cm$^{-2}$ (vertical gas column density of 2$\times$10$^{23}$~cm$^{-2}$),
the value at the outer radius of the \citeauthor{pinte} model. We have
run a number of disc models, with the inner 400~AU described by the
\citeauthor{pinte} model (with the gas kinetic temperature scaled as
described in the previous paragraph) and the region from 400 to 900~AU
described here by the disc extension. Figure~\ref{plplots} shows the surface density in the models that we have tested: within 400~AU it is the surface density as in \citeauthor{pinte} (blue line) and between 400 and 900~AU different combinations of $\Sigma_{400}$ and $p$ (black lines).
The models are tested against the observed $^{12}$CO and $^{13}$CO \textit{uv}-data, channel
maps, spectra, and position-velocity plots. The comparison of modelled emission with \textit{uv}-data for the line wings, $v_{\rm LSR}<$3.0~km~s$^{-1}$ for $^{12}$CO and $v_{\rm LSR}<$3.5~km~s$^{-1}$+$v_{\rm LSR}>$5.5~km~s$^{-1}$ for $^{13}$CO is also examined. 

Figure~\ref{plplots} shows the models that overproduce the observed emission with dashed black lines and those that underproduce it with dotted black lines. The full black lines correspond to the models that reproduce well our $^{12}$CO and $^{13}$CO data. The general area (beyond 400~AU) allowed by the models is shaded in Fig.~\ref{plplots} for guidance. It can be seen that the $^{12}$CO and $^{13}$CO
observations constrain the column density of $^{12}$CO at $R=900$~AU
to $N_{\rm CO}=(0.05-1.0)\times10^{18}$~cm$^{-2}$,
where the lower bound follows from the requirement that the $^{12}$CO emission is sufficently extended and the upper bound from the requirement that the $^{12}$CO and $^{13}$CO peak intensity, and the 
extent of the $^{13}$CO emission are not overestimated.
The corresponding surface density at 900~AU is
$\Sigma_{900}=(0.2-4.0)\times 10^{-2}$~g~cm$^{-2}$, i.e.,
a vertical gas column density $(0.05$--$1.0)\times 10^{22}$~cm$^{-2}$.
Our data do not constrain the parameters $\Sigma_{400}$ and $p$, that determine how the surface density decreases from its value at the outer edge of the \citeauthor{pinte} model, to its value at 900~AU. This is either a marked change from the power-law slope of $p=1$ found
inside 400~AU to $p=5$ beyond 400~AU, or a discontinuous drop by a factor $\sim$10-100 in surface density at 400~AU. 

\begin{figure}
\includegraphics[angle=0,width=8cm]{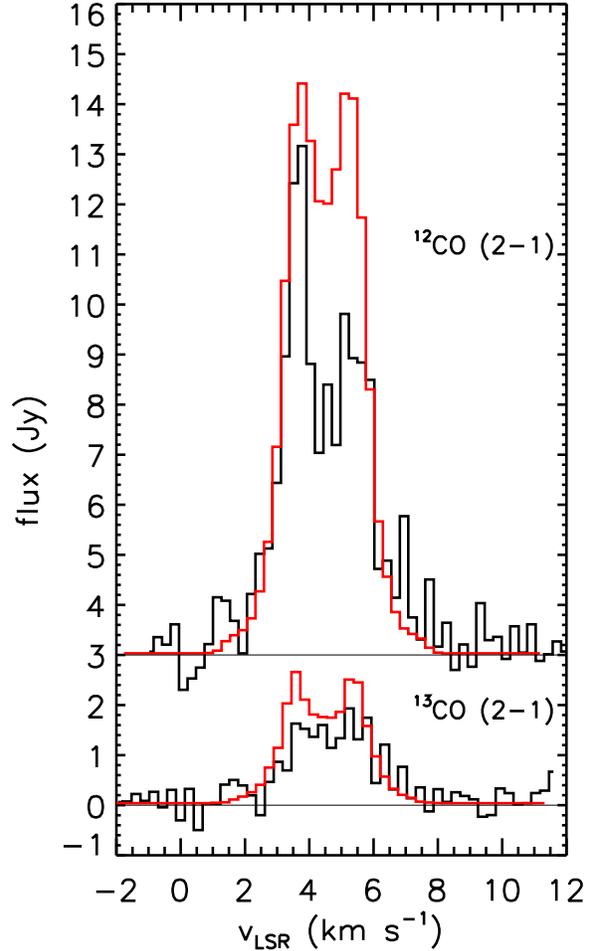}
\caption
      {$^{12}$CO and $^{13}$CO $J$=2--1 line spectra averaged over a
        $8\farcs0\times8\farcs0$ region centered on IM Lup. The
        $^{12}$CO spectrum is shifted up by 3~Jy for clarity. The red
        lines show the emission predicted the extended disc model described in
        Sect.~\ref{s:model} with $\Sigma_{400}=2\times10^{21}$cm$^{-2}$ and $p=1$.}
\label{4specA}
\end{figure}

\begin{figure}
\includegraphics[angle=0,width=9cm]{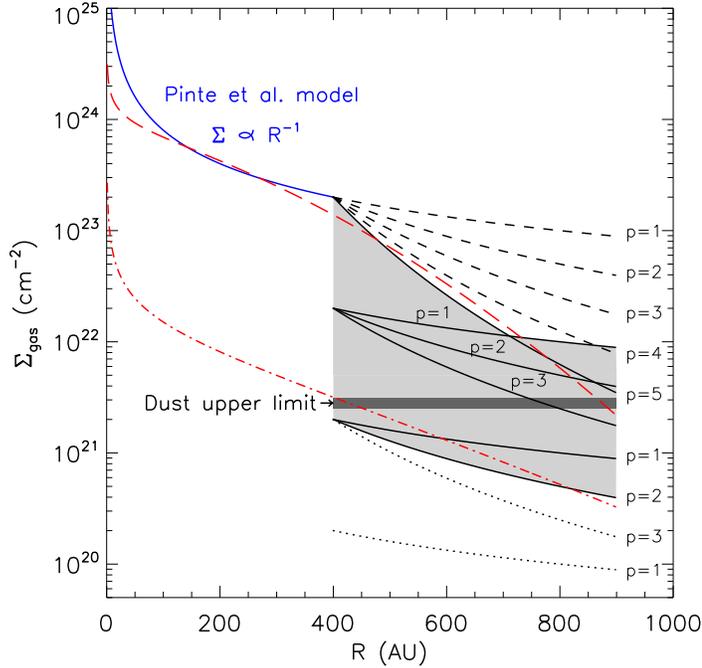}
\caption
      {Gas surface density in our disc models is plotted as a function of radius. Within 400~AU, it is identical to the Pinte et al. model shown with the full blue line. Outside 400~AU, we explore different power-law distributions, each plotted in black and marked with the corresponding slope $p$. The models which overestimate the observed $^{12}$CO emission are plotted with dashed lines, while those that underpredict it are shown in dotted lines.The full black lines represent the models that fit well the $^{12}$CO  $J$=2--1 emission, and define the shaded region which shows our constraint on $\Sigma_{\rm CO}/$[CO] in the outer disc, where a CO abundance of [CO]=10$^{-4}$ is used. The upper limit on $\Sigma_{\rm dust}\,g/d$ placed by scattered light images is shown with a dark grey line, with a gas-to-dust mass ratio $g/d$=100. For comparison, the red lines correspond to disc models with an exponential drop off as described in \citet{mhughes}. The model shown with the long-dashed red line has parameters $\gamma=$0.3, $c_1$=340~AU and $c_2=3.1\times10^{24}$~cm$^{-2}$ and fits the gas constraints. The model with the dash-dotted red line  $\gamma=$0.6, $c_1$=340~AU and $c_2=1.8\times10^{23}$~cm$^{-2}$ fits the scattered light constraints. No single model with a tapered outer edge can fit both these constraints and the constraints within 400~AU simultaneously.} 
\label{plplots}
\end{figure}

Figure~\ref{uvco} compares observations to synthetic
$^{12}$CO and $^{13}$CO line visibilities for our model
with $\Sigma_{400}=2\times10^{21}$cm$^{-2}$ and
$p=1$, plotted with dotted red lines. There is a good match between the model and the data for both
transitions. In particular, the model
reproduces well the change in the slope of visibilities, mentioned in Sect.~\ref{s:lines}. 
The match between the model (red lines) and observations is also seen in the line spectra, Fig.~\ref{4specA}.
Figures~\ref{maps12} and \ref{mapsA13} show, respectively, the
$^{12}$CO and $^{13}$CO channel maps (black contours) compared to our extended disc model (grey contours). 
It can therefore be seen that our model provides a good description not only of the line intensity at each channel (spectra), but also a very close match in the spatial extent of the emission in each spectral channel.

A good fit to the wings of the $^{12}$CO and $^{13}$CO spectra (Fig.~\ref{4specA}, red lines) and the spatial distribution of the respective line fluxes at highly blue- and red-shifted velocities (lower panels, Figs.\ref{maps12} and \ref{mapsA13})
is found for
temperature scalings $f$ of 1.7 for $^{12}$CO and 1.4 for $^{13}$CO. These values of $f$ suggest that the
gas is somewhat warmer than the dust at the heights above the disc where the
$^{12}$CO and $^{13}$CO emission originates, and more so at the larger height probed by the $^{12}$CO line compared to the $^{13}$CO line.

At the adopted disc inclination of $50\degr$, the line peak separation provides
a reliable contraint on the stellar mass. We find a best-fit of
$M_\star= 1.2 \pm 0.4$~M$_\odot$, where the uncertainty is dominated by
our limited spectral resolution. This value is consistent with the rough estimate of 1~M$_{\odot}$
from \citet{pinte}, but a few times higher than derived by \citet{hughes}.

We conclude that the surface density
\emph{traced through $^{12}${\rm CO} and $^{13}${\rm CO}} has a discontinuity
around $R=$400~AU either in $\Sigma_{\rm CO}(R)$ or in its derivative
$d\Sigma_{\rm CO}/dR$, or both. This may, or may not be an indication of an overall 
discontinuity of the gas surface density.
A break in the temperature $T(R)$ cannot
explain the observations, since our model already adopts a low
temperature at the margin of $^{12}$CO freeze-out in the outer regions.
An alternative explanation for the observations is a radical drop in the
abundance of CO (with respect to H$_2$ and H) or its radial derivative. Freeze out onto dust grains or
photodissociation can significantly reduce the gas-phase abundance of
CO. In the next section we explore the limits that the dust emission
can give us on the gas content outside 400~AU, and compare them to the $^{12}$CO results.

\subsection{Comparing gas and dust at radii beyond 400 AU}\label{s:cfdust}

The previous section concluded that both the gas and the dust out to 400~AU in the
disc around IM~Lup is well described by the model of
\citeauthor{pinte}, with the exception of gas
temperatures that exceed the dust temperature at some height above the
disc midplane. It also found that the \emph{gas} disc needs to be
extended to an outer radius of 900~AU, albeit with a significant
decrease in the surface density of CO, $\Sigma_{\rm CO}$, or in its
first derivative, $d\Sigma_{\rm CO}/dR$ close to 400~AU. 

\citet{pinte} show that
some dust is present outside 400~AU as well, visible as an extended
nebulosity in their 0.8~$\mu$m scattered light images. At the same time, the visible \emph{lower} scattering disc surface places a stringent limit on the surface density $\Sigma_{\rm dust}$ of small dust
particles outside 400~AU.
Requiring the optical depth {$\tau=\Sigma_{dust}\,\kappa< 1$}
and adopting an emissivity per gram of dust of
{$\kappa$=(8-10)$\times$10$^{3}$~cm$^2$g$^{-1}$} at 0.8~$\mu$m \citep[See first row of Tab.~1,][]{ossenkopf}, we find {$\Sigma_{dust}\leq (1.0-1.3)\times
10^{-4}$~g~cm$^{-2}$}. If we adopt the gas-to-dust mass ratio of 100, this corresponds to {$N_{\rm H_2}\leq (2.5-3.1)\times10^{21}$~cm$^{-2}$}. 
Our limit differs from that given in \citet{pinte} (0.2~g~cm$^{-2}$) because we use dust opacities representative of small dust, while they assume considerable grain growth in disc midplane and thus use much lower dust opacities at 0.8~$\mu$m.
The limit on surface density we derive is two orders of magnitude lower than the column density at the
outer radius of 400~AU of the \citeauthor{pinte} model. This indicates that either the dust surface density drops sharply at 400~AU, or that efficient grain growth beyond 400~AU has caused a significant decrease in dust near-IR opacity.
As can be seen in Fig.~\ref{plplots}, the upper limit on surface density of $(2.5-3.1)\times
10^{21}$~cm$^{-2}$ is consistent with the gas surface density range inferred in Sect.~\ref{s:model} from our
CO data, using the canonical CO/H$_2$ abundance of 10$^{-4}$.

While 0.8~$\mu$m imaging traces the small dust, our observations of 1.3~mm dust continuum emission, on the other hand, trace the millimetre-sized dust particles. In Fig.~\ref{uvamp} we can see that the \citeauthor{pinte} model (full red line), with the radius of 400~AU, compares well to the observed continuum flux at all projected baseline lengths. On the other hand, the comparison of the 1.3~mm visibilities to our extended disc model with $\Sigma_{400}$=2$\times10^{21}$~cm$^{-2}$ and $p=$1 shows that the model overestimates emission at short \textit{uv}-distances (large spatial scales). A constant dust emissivity of 2.0~cm$^2$~g$^{-1}$ \citep[emissivity of mm-sized grains, as in][]{draine} was used throughout the disc in the calculation of 1.3~mm fluxes. Our model indicates that any dust present in the outer disc regions must be poor in mm-sized grains, i.e., have low millimetre wavelength opacities, while dust within 400~AU has likely undergone grain growth as found by \citet{pinte}. This further supports our choice of $\kappa$ at 0.8~$\mu$m when estimating the upper limit on dust column (see above). 
Therefore, a viable model for the disc of IM~Lup consists of an
`inner' disc extending to 400~AU as described in \citeauthor{pinte}
augmented with an `outer disc' extending from 400 to 900~AU with a
significantly reduced surface density (with negligible mass) but standard
gas-to-dust mass ratio and CO-to-H$_2$ ratios. The SED of this new model should not differ significantly to that of \citeauthor{pinte} model, and is therefore expected to provide a good match to the observed SED of IM Lup.

\citet{mhughes} find that the apparent difference between the extent of submillimetre dust and gas emission in several circumstellar discs can be explained by an exponential drop off of surface density which naturally arises at the outer edge of a viscous disc. In Fig.~\ref{plplots} we show how, with a careful choice of parameters ($\gamma=$0.3, $c_1$=340~AU and $c_2=3.1\times10^{24}$~cm$^{-2}$), the model of \citet{mhughes} (red long-dashed line) can reproduce the surface density distribution of the models which describe well the $^{12}$CO 2--1 line emission. This model, and the one discussed below, are only examples. A proper modelling of IM Lup in the context of viscous disc models would require a revision of the entire disc structure both in terms of temperature and density, which is outside of the scope of the current work. We notice that the \citeauthor{mhughes} models cannot simultaneously comply with the gas and dust constraints in the outer disc and the surface density derived by \citet{pinte} in the inner disc. This is illustrated by the \citet{mhughes} model with parameters $\gamma=$0.6, $c_1$=340~AU and $c_2=1.8\times10^{23}$~cm$^{-2}$, shown with the red dash-dotted line in Fig.~\ref{plplots}. The surface density of this model outside 400~AU is in agreement with observational constraints from gas and dust, but it is roughly two orders of magnitude lower than suface density from \citet{pinte} within 400~AU.

In the standard theory of viscous discs \citep[See][]{pringle},
irrespective of the initial density distribution, a radially smooth
surface density distribution with a tapered outer edge is
rapidly reached. If there is a significant change in the nature of the
viscosity inside and outside of 400~AU, discontinuities in the
equilibrium surface density may follow. Such changes could, for
example, result from differences in the ionization structure of the
disc or from a drop of the surface density below some critical
level. 
Here we explore some scenarios that could explain this:

\textit{A young disc.}
An extreme example of such a configuration is a disc where the inner
400~AU follows the standard picture of a viscous accretion disc, but
where the region outside 400~AU has not (yet) interacted viscously
with the inner disc. This outer region may be the remnant of the
flattened, rotating prestellar core that has not yet made it onto the
viscous inner disc. This configuration, reminiscent of the material
around the object L1489~IRS \citep{brinch},
suggests that IM~Lup would only recently have emerged from the
embedded phase.
L1489~IRS showed clear inward motion in its rotating envelope. Our
observations limit any radial motions in the gas between 400 and
900~AU to $<0.2$~km~s$^{-1}$, or 20\% of the Keplerian orbital
velocities at these radii. Furthermore, for the 900~AU structure to
survive for the lifetime of IM~Lup of 0.1--0.6~Myr, inward motions cannot
exceed $\sim$10$^{-2}$~km~s$^{-1}$. Any mass inflow is therefore small, and the
material between 400 and 900~AU is likely on Keplerian orbits.

\textit{A companion body.}
Another explanation for the break in the disc density structure around
400~AU would be the presence of a companion near this radius. A companion
of $\sim 1$~M$_{\rm Jup}$ at 400~AU could open a gap in the disc and
affect the viscous disc spreading. No companions at this separation
are visible in the HST images of \citep{pinte} or in $K$-band direct
imaging \citep{ghez}. Whether these observations exclude this
scenario is unclear: it requires modelling of the orbital evolution
of a companion in a viscously spreading disc and calculation of the
observational mass limits at the age of IM~Lup. This is beyond the
scope of this paper.

\textit{Gas to dust ratio.}
While our model is consistent with standard gas-to-dust and
CO-to-H$_2$ ratios beyond 400~AU, this is not the only
solution. Instead of adopting these standard ratios, which requires
explaining the drop in $\Sigma$ or $d\Sigma/dR$ around 400~AU, we can
hypothesize that the gas (H$_2$ or H) surface density is continuous
out to 900~AU and that both the CO-to-H$_2$ and dust-to-gas ratios
show a break around 400~AU. This scenario requires a drop between 400
and 900 AU of the CO abundance by a factor between 10 and 200, and of the
dust-to-gas mass ratio by a factor $\ge$90. These drops can be sudden,
with a discontinuity at 400~AU, or more gradual, with a rapid decline
of the two ratios from 400 to 900~AU. Since a low amount of dust
emission outside 400~AU is observed both at wavelengths of $\sim 1$~mm
(our data) and $\sim 1$~$\mu$m \citep{pinte}, the overall dust-to-gas
ratio is likely affected, and not just the individual populations of
small and large grains.

\textit{Dust radial drift and photoevaporation.}
If a large fraction of the dust is removed from the disc regions
outside 400~AU, the increased penetration of ultraviolet radiation
could explain the drop in $^{12}$CO surface density through increased
photodissociation \citep{zadelhoff}. Radial drift of dust particles due to the gas drag force \citep{whipple, weidenschilling} is a possible scenario in circumstellar discs. The difference in velocity between the dust, in Keplerian rotation, and gas, sub-Keplerian because of the radial pressure gradient, can cause dust particles to lose angular momentum and drift inward. The optimal drift particle size depends on the gas density, Keplerian rotation frequency and hydrostatic sound speed.  
Most dust evolution models focus on the inner 100~AU of discs, relevant to planet
formation. In these regions, the grains from 100~$\mu$m to about 0.1~m
efficiently migrate inwards on a timescale shorter than 2~Myr. However, the optimal grain size for
inward drift decreases with the gas density. Our modelling of the disc region from 400~AU to 900~AU, predicts
surface densities of $\sim 10^{-3}$~g~cm$^{-2}$, low enough even for
sub-micron-sized particles to drift inward (to $<400$~AU). For the
estimated age of IM~Lup of 0.1--0.6~Myr, all particles larger than
0.1--0.02~$\mu$m will have migrated inward. 
This process leaves the outer disc unshielded by dust against UV radiation. Infrared emission of PAHs may be used to trace the disc surface in this scenario. However, \citet{geers} do not detect PAH emission at 3.3~$\mu$m in their VLT-ISAAC L-band observations of IM Lup. This may indicate that either there are not enough PAHs in the disc or that they are not exposed to a significant level of UV flux. The latter possibility allows the outer disc to remain molecular. Otherwise, the outer disc is exposed to photodissociating radiation, destroying much of the CO and likely also a significant
fraction of the H$_2$ given the limit on the dust surface density of
$10^{21}$~cm$^{-2}$ corresponding to $A_V\approx 1^{\rm mag}$.  In
this scenario, the outer disc between 400 and 900~AU would be largely
\emph{atomic} and possibly detectable through 21~cm observations of
H~I, or line observations of C~I at 609 and 370~$\mu$m or C~II at
158~$\mu$m. If \emph{photoevaporation} is efficient in this region
it may remove the (atomic) gas and reduce the gas surface density further.
Therefore, a combined effect of efficient drift, photodissociation and photoevaporation in the outermost disc regions may be a reason for the low gas and dust density observed. The efficiency of these processes decreases with density and perhaps the density at 400~AU is high enough so that material is no longer efficiently removed from the disc. Only the detailed simultaneous modelling of drift, photodissociation and photoevaporation could test this scenario.


\section{Conclusions}\label{s:conclusions}

We probe the kinematics and the distribution of the gas and dust in
the disc around IM~Lup through molecular gas and continuum dust
emission. Our SMA observations resolve the disc structure down to
scales of 200~AU, and allow us to probe the structure of the inner
disc ($<400$~AU) and the outer disc (400--900~AU). Our main
conclusions can be summarized as follows.

\begin{itemize}
\item The $^{12}$CO and $^{13}$CO emission extends to 900~AU from
  IM~Lup, much further than the outer radius of 400~AU inferred
  earlier from dust measurements.

\item The H$_2$ gas surface density in the region between 400 and
  900~AU lies in the range of $5 \times 10^{20}$ to $10
  ^{22}$~cm$^{-2}$, using the standard CO-to-H$_2$ ratio of 10$^{-4}$.

\item The disc is in Keplerian rotation around a central mass of $1.2
  \pm 0.4$~M$_\odot$. Infall motions, if present in the outer disc, are
  negligible at $<0.2$~km~s$^{-1}$.

\item The molecular line emission from the inner disc, within 400~AU,
  is well described by the model of \citet{pinte}, except that the gas
  temperature in the layers dominating the line emission of $^{12}$CO
  and $^{13}$CO exceeds the dust temperature by factors 1.7 and 1.4,
  respectively. 

\item Outside 400~AU, the surface densities of the molecular gas, as
  traced through $^{12}$CO and $^{13}$CO, of small ($\sim 1$~$\mu$m)
  dust grains, and of larger ($\sim 1$~mm) dust grains have a break in
  their radial dependence. At 400~AU, the dust surface density (in
  small grains) drops by a factor $\sim$100, while the gas surface density
  shows a comparable drop of a factor 10--200 or steepens its radial
  power-law slope from $p=1$ to $5\leq p\leq 8$.

\end{itemize}

Our observations show that the disc around IM~Lup consists of two
regions. The inner 400~AU is well described by a `standard' accretion
disc; the region between 400 and 900~AU has a much lower surface
density as traced through dust grains with sizes from $\sim 1$~$\mu$m
to 1~mm and through CO emission. Our observations do not tell us if
this outer region consists of material from the original prestellar
core that has not (yet) made it onto the viscous accretion disc, or of
material that is part of the disc but has had a different
evolution. Sensitive, spatially resolved observations at various (sub)
millimetre wavelengths, as may be obtained with the Atacama Large
Millimeter Array may help to assess whether
significantly different grain populations exist inside and outside of
400~AU. With the same telescope, very high signal-to-noise
observations of $^{12}$CO lines at high spectral resolution may allow
determination of any radial (inward or outward) motions in the
$>400$~AU gas. Spatially resolved mid-infrared imaging in several
emission bands of PAHs, as could be obtained with the VISIR instrument
on VLT, would shed light on the question if the 400--900~AU zone in
the disc is largely photodissociated or -ionized. Detailed modelling of dust evolution in the outer disc may answer whether radial drift is responsible for the low column of dust beyond 400~AU in IM Lup disc.

\begin{acknowledgements}
The research of O.~P. and M.~R.~H. is supported through a VIDI grant
from the Netherlands Organisation for Scientific Research. We would like to thank our Leiden colleagues Anders Johansen and Richard D. Alexander for valuable insights and discussions, as well as C.~P. Dullemond, A. Juh\'asz and others at the Star and Planet Formation Department of MPIA Heidelberg for their help and advice during the stay of O.~P. in March 2008. Finally, we are grateful to E.~F. van Dishoeck for her support and guidance throughout the writing of this paper.
\end{acknowledgements}

\bibliographystyle{aa}
\bibliography{/data1/olja/tex/imluptex/refs}

\begin{thebibliography}{36}
\expandafter\ifx\csname natexlab\endcsname\relax\def\natexlab#1{#1}\fi

\bibitem[{{Baraffe} {et~al.}(1998){Baraffe}, {Chabrier}, {Allard}, \&
  {Hauschildt}}]{baraffe}
{Baraffe}, I., {Chabrier}, G., {Allard}, F., \& {Hauschildt}, P. 1998, VizieR
  Online Data Catalog, 333, 70403

\bibitem[{{Brinch} {et~al.}(2007){Brinch}, {Crapsi}, {Hogerheijde}, \&
  {J{\o}rgensen}}]{brinch}
{Brinch}, C., {Crapsi}, A., {Hogerheijde}, M.~R., \& {J{\o}rgensen}, J.~K.
  2007, \aap, 461, 1037

\bibitem[{{Chiang} \& {Goldreich}(1997)}]{chiang}
{Chiang}, E.~I. \& {Goldreich}, P. 1997, \apj, 490, 368

\bibitem[{{D'Alessio} {et~al.}(2005){D'Alessio}, {Mer{\'{\i}}n}, {Calvet},
  {Hartmann}, \& {Montesinos}}]{dalessio}
{D'Alessio}, P., {Mer{\'{\i}}n}, B., {Calvet}, N., {Hartmann}, L., \&
  {Montesinos}, B. 2005, Revista Mexicana de Astronomia y Astrofisica, 41, 61

\bibitem[{{D'Antona} \& {Mazzitelli}(1994)}]{d'antona94}
{D'Antona}, F. \& {Mazzitelli}, I. 1994, \apjs, 90, 467

\bibitem[{{Dartois} {et~al.}(2003){Dartois}, {Dutrey}, \&
  {Guilloteau}}]{dartois}
{Dartois}, E., {Dutrey}, A., \& {Guilloteau}, S. 2003, \aap, 399, 773

\bibitem[{{Draine}(2006)}]{draine}
{Draine}, B.~T. 2006, \apj, 636, 1114

\bibitem[{{Dullemond} {et~al.}(2001){Dullemond}, {Dominik}, \&
  {Natta}}]{dullemond}
{Dullemond}, C.~P., {Dominik}, C., \& {Natta}, A. 2001, \apj, 560, 957

\bibitem[{{Frerking} {et~al.}(1982){Frerking}, {Langer}, \&
  {Wilson}}]{frerking}
{Frerking}, M.~A., {Langer}, W.~D., \& {Wilson}, R.~W. 1982, \apj, 262, 590

\bibitem[{{Geers} {et~al.}(2007){Geers}, {van Dishoeck}, {Visser},
  {Pontoppidan}, {Augereau}, {Habart}, \& {Lagrange}}]{geers}
{Geers}, V.~C., {van Dishoeck}, E.~F., {Visser}, R., {et~al.} 2007, \aap, 476,
  279

\bibitem[{{Ghez} {et~al.}(1997){Ghez}, {McCarthy}, {Patience}, \&
  {Beck}}]{ghez}
{Ghez}, A.~M., {McCarthy}, D.~W., {Patience}, J.~L., \& {Beck}, T.~L. 1997,
  \apj, 481, 378

\bibitem[{{Guilloteau} \& {Dutrey}(1998)}]{guilloteau}
{Guilloteau}, S. \& {Dutrey}, A. 1998, \aap, 339, 467

\bibitem[{{Hogerheijde} \& {van der Tak}(2000)}]{hogerheijde}
{Hogerheijde}, M.~R. \& {van der Tak}, F.~F.~S. 2000, \aap, 362, 697

\bibitem[{{Hughes} {et~al.}(2008){Hughes}, {Wilner}, {Qi}, \&
  {Hogerheijde}}]{mhughes}
{Hughes}, A.~M., {Wilner}, D.~J., {Qi}, C., \& {Hogerheijde}, M.~R. 2008, \apj,
  678, 1119

\bibitem[{{Hughes} {et~al.}(1994){Hughes}, {Hartigan}, {Krautter}, \&
  {Kelemen}}]{hughes}
{Hughes}, J., {Hartigan}, P., {Krautter}, J., \& {Kelemen}, J. 1994, \aj, 108,
  1071

\bibitem[{{Isella} {et~al.}(2007){Isella}, {Testi}, {Natta}, {Neri}, {Wilner},
  \& {Qi}}]{isella}
{Isella}, A., {Testi}, L., {Natta}, A., {et~al.} 2007, \aap, 469, 213

\bibitem[{{Jonkheid} {et~al.}(2004){Jonkheid}, {Faas}, {van Zadelhoff}, \& {van
  Dishoeck}}]{jonkheid}
{Jonkheid}, B., {Faas}, F.~G.~A., {van Zadelhoff}, G.-J., \& {van Dishoeck},
  E.~F. 2004, \aap, 428, 511

\bibitem[{{J{\o}rgensen} {et~al.}(2005){J{\o}rgensen}, {Bourke}, {Myers},
  {Sch{\"o}ier}, {van Dishoeck}, \& {Wilner}}]{jorgensen}
{J{\o}rgensen}, J.~K., {Bourke}, T.~L., {Myers}, P.~C., {et~al.} 2005, \apj,
  632, 973

\bibitem[{{Lommen} {et~al.}(2007){Lommen}, {Wright}, {Maddison},
  {J{\o}rgensen}, {Bourke}, {van Dishoeck}, {Hughes}, {Wilner}, {Burton}, \&
  {van Langevelde}}]{lommen}
{Lommen}, D., {Wright}, C.~M., {Maddison}, S.~T., {et~al.} 2007, \aap, 462, 211

\bibitem[{{Ossenkopf} \& {Henning}(1994)}]{ossenkopf}
{Ossenkopf}, V. \& {Henning}, T. 1994, \aap, 291, 943

\bibitem[{{Pani{\'c}} {et~al.}(2008){Pani{\'c}}, {Hogerheijde}, {Wilner}, \&
  {Qi}}]{panic}
{Pani{\'c}}, O., {Hogerheijde}, M.~R., {Wilner}, D., \& {Qi}, C. 2008, ArXiv
  e-prints

\bibitem[{{Pi{\'e}tu} {et~al.}(2007){Pi{\'e}tu}, {Dutrey}, \&
  {Guilloteau}}]{pietu}
{Pi{\'e}tu}, V., {Dutrey}, A., \& {Guilloteau}, S. 2007, \aap, 467, 163

\bibitem[{{Pinte} {et~al.}(2008){Pinte}, {Padgett}, {Menard}, {Stapelfeldt},
  {Schneider}, {Olofsson}, {Panic}, {Augereau}, {Duchene}, {Krist},
  {Pontoppidan}, {Perrin}, {Grady}, {Kessler-Silacci}, {van Dishoeck},
  {Lommen}, {Silverstone}, {Hines}, {Wolf}, {Blake}, {Henning}, \&
  {Stecklum}}]{pinte}
{Pinte}, C., {Padgett}, D.~L., {Menard}, F., {et~al.} 2008, ArXiv e-prints, 808

\bibitem[{{Pringle}(1981)}]{pringle}
{Pringle}, J.~E. 1981, \araa, 19, 137

\bibitem[{{Qi} {et~al.}(2004){Qi}, {Ho}, {Wilner}, {Takakuwa}, {Hirano},
  {Ohashi}, {Bourke}, {Zhang}, {Blake}, {Hogerheijde}, {Saito}, {Choi}, \&
  {Yang}}]{qi}
{Qi}, C., {Ho}, P.~T.~P., {Wilner}, D.~J., {et~al.} 2004, \apjl, 616, L11

\bibitem[{{Qi} {et~al.}(2006){Qi}, {Wilner}, {Calvet}, {Bourke}, {Blake},
  {Hogerheijde}, {Ho}, \& {Bergin}}]{qi1}
{Qi}, C., {Wilner}, D.~J., {Calvet}, N., {et~al.} 2006, \apjl, 636, L157

\bibitem[{{Raman} {et~al.}(2006){Raman}, {Lisanti}, {Wilner}, {Qi}, \&
  {Hogerheijde}}]{raman}
{Raman}, A., {Lisanti}, M., {Wilner}, D.~J., {Qi}, C., \& {Hogerheijde}, M.
  2006, \aj, 131, 2290

\bibitem[{{Sault} {et~al.}(1995){Sault}, {Teuben}, \& {Wright}}]{sault}
{Sault}, R.~J., {Teuben}, P.~J., \& {Wright}, M.~C.~H. 1995, in Astronomical
  Society of the Pacific Conference Series, Vol.~77, Astronomical Data Analysis
  Software and Systems IV, ed. R.~A. {Shaw}, H.~E. {Payne}, \& J.~J.~E.
  {Hayes}, 433--+

\bibitem[{{Sch{\"o}ier} {et~al.}(2005){Sch{\"o}ier}, {van der Tak}, {van
  Dishoeck}, \& {Black}}]{schoier}
{Sch{\"o}ier}, F.~L., {van der Tak}, F.~F.~S., {van Dishoeck}, E.~F., \&
  {Black}, J.~H. 2005, \aap, 432, 369

\bibitem[{{Swenson} {et~al.}(1994){Swenson}, {Faulkner}, {Rogers}, \&
  {Iglesias}}]{swenson94}
{Swenson}, F.~J., {Faulkner}, J., {Rogers}, F.~J., \& {Iglesias}, C.~A. 1994,
  \apj, 425, 286

\bibitem[{{van Kempen} {et~al.}(2007){van Kempen}, {van Dishoeck}, {Brinch}, \&
  {Hogerheijde}}]{vankempen}
{van Kempen}, T.~A., {van Dishoeck}, E.~F., {Brinch}, C., \& {Hogerheijde},
  M.~R. 2007, \aap, 461, 983

\bibitem[{{van Zadelhoff} {et~al.}(2003){van Zadelhoff}, {Aikawa},
  {Hogerheijde}, \& {van Dishoeck}}]{zadelhoff}
{van Zadelhoff}, G.-J., {Aikawa}, Y., {Hogerheijde}, M.~R., \& {van Dishoeck},
  E.~F. 2003, \aap, 397, 789

\bibitem[{{Weidenschilling}(1977)}]{weidenschilling}
{Weidenschilling}, S.~J. 1977, \mnras, 180, 57

\bibitem[{{Whipple}(1972)}]{whipple}
{Whipple}, F.~L. 1972, in From Plasma to Planet, ed. A.~Elvius

\bibitem[{{Wichmann} {et~al.}(1998){Wichmann}, {Bastian}, {Krautter},
  {Jankovics}, \& {Rucinski}}]{wichmann}
{Wichmann}, R., {Bastian}, U., {Krautter}, J., {Jankovics}, I., \& {Rucinski},
  S.~M. 1998, \mnras, 301, L39+

\bibitem[{{Wilson} \& {Rood}(1994)}]{wilson}
{Wilson}, T.~L. \& {Rood}, R. 1994, \araa, 32, 191

\end{thebibliography}
\end{document}